\documentclass[prb,reprint,superscriptaddress,citeautoscript]{revtex4-1}
\newcommand{\headline}[1]{ {\it #1 -- }}

\usepackage[applemac]{inputenc}
\usepackage[T1]{fontenc}
\usepackage[american]{babel}
\usepackage[scaled]{helvet}
\usepackage{graphicx,amssymb,bm,mathtools,textcomp,courier,multirow,hyperref,color,pdfpages,subfigure}
\usepackage[normalem]{ulem}	%fixes problem with book references

\newcommand{\op}[2]{\left.{\left| #1\right\rangle\left\langle #2\right|}\right.}

\newcommand{\ket}[1]{\left|\left. #1 \right\rangle\right.}

\newcommand{\figureshortname}{Fig.}
\newcommand{\equationshortname}{Eq.}

\newcommand{\meref}[1]{Eqs.~\eqref{#1}}
\newcommand{\eref}[1]{\equationshortname~\eqref{#1}}
\newcommand{\sref}[1]{Sec.~\ref{#1}}
\newcommand{\aref}[1]{Appx.~\ref{#1}}
\newcommand{\fref}[1]{\figureshortname~\ref{#1}}

\newcommand{\rcite}[1]{Ref.~[\onlinecite{#1}]}

\newcommand\undermat[2]{%
  \makebox[0pt][l]{$\smash{\underbrace{
  \phantom{\begin{matrix}#2\end{matrix}}}
  _{\text{$#1$}}}$}#2}

\graphicspath{{Pictures/}}

\begin{document}
\def\sectionautorefname{Sec.}

\title{Two-Qubit Couplings of Singlet-Triplet Qubits Mediated by One Quantum State}

\author{Sebastian Mehl}
\email{s.mehl@fz-juelich.de}
\affiliation{Peter Grünberg Institute (PGI-2), Forschungszentrum Jülich, D-52425 Jülich, Germany}
\affiliation{JARA-Institute for Quantum Information, RWTH Aachen University, D-52056 Aachen, Germany}
\author{Hendrik Bluhm}
\affiliation{JARA-Institute for Quantum Information, RWTH Aachen University, D-52056 Aachen, Germany}
\author{David P. DiVincenzo}
\affiliation{Peter Grünberg Institute (PGI-2), Forschungszentrum Jülich, D-52425 Jülich, Germany}
\affiliation{JARA-Institute for Quantum Information, RWTH Aachen University, D-52056 Aachen, Germany}

\date{\today}

%------------------------------------------------------------------------------------
%------------------------------------------------------------------------------------
%------------------------------------------------------------------------------------
\begin{abstract}
We describe high-fidelity entangling gates between singlet-triplet qubits (STQs) which are coupled via one quantum state (QS). The QS can be provided by a quantum dot itself or by another confined system. The orbital energies of the QS are tunable using an electric gate close to the QS, which changes the interactions between the STQs independent of their single-qubit parameters. Short gating sequences exist for the controlled $\text{NOT}$ ($\text{CNOT}$) operations. We show that realistic quantum dot setups permit excellent entangling operations with gate infidelities below $10^{-3}$, which is lower than the quantum error correction threshold of the surface code. We consider limitations from fabrication errors, hyperfine interactions, spin-orbit interactions, and charge noise in GaAs and Si heterostructures.
\end{abstract}

\maketitle

%-----------------------------------------------------------------------------------
%-----------------------------------------------------------------------------------
%-----------------------------------------------------------------------------------
\section{Introduction}
A spin-based quantum computer can be realized using singlet-triplet qubits (STQs) \cite{levy2002,taylor2005,taylor2007}. One qubit is encoded in the ${s_z=0}$ spin subspace of two singly occupied quantum dots (QDs). Single-qubit control is provided by the exchange interaction between the electrons on the two QDs \cite{petta2005,johnson2005} and a magnetic field gradient over the double quantum dot (DQD) \cite{foletti2009,bluhm2010,gullans2010,tokura2006,pioro-ladriere2008,brunner2011}. The magnitude of the exchange interaction can be tuned rapidly using electric gates near the QDs.  
Single-qubit control of a STQ is extremely successful for gate-defined QDs in GaAs \cite{foletti2009,brunner2011} and Si \cite{maune2012}; low-frequency noise is successfully eliminated in decoupling experiments \cite{barthel2010-1,bluhm2011}.

Two-qubit gates are more demanding for STQs. Two approaches have been suggested. Electrostatic couplings between STQs provide two-qubit interactions \cite{taylor2005,hanson2007-1}. When a DQD is biased using electric fields, only the singlet state allows the transfer of one electron to the doubly occupied configuration on one QD. The charge configurations of the singlet and the triplet states differ for a biased DQD. Coulomb interactions create an energy shift for one STQ conditioned on the state of the other STQ \cite{taylor2005,hanson2007-1}. A controlled phase gate was demonstrated experimentally \cite{shulman2012}. However, electrostatic couplings are usually weak, which makes these operations slow. Alternatively, direct exchange interactions between the DQDs can be used. This approach was originally introduced for single-electron spin qubits \cite{loss1998}. The realization of direct exchange gates between STQs has not been successful so far. The DQDs must be close to each other to allow an overlap of the electrons' wave functions. Note that optical manipulations of QDs provide additional possibilities for entangling operations. A two-qubit gate with $80\%$ fidelity was demonstrated using laser driving to an excited quantum state (QS) \cite{kim2011}.

In this paper we explore indirect exchange interactions between STQs via one QS. This approach was already proposed in passing in \rcite{loss1998}. We explore the rich opportunities of mediated couplings while considering all possible charge configurations of the QS. The QS can be empty, singly occupied, or filled with two electrons. Each charge configuration permits entangling operations for STQs. We describe entangling gate sequences which are shorter than all earlier proposals for direct exchange interactions \cite{levy2002,klinovaja2012} and do not require the interaction strength to be raised to unrealistically large values \cite{levy2002}. Our gate sequences are high fidelity even without applying complicated noise corrections \cite{kestner2013}. 
Gate infidelities below $10^{-3}$ can be realized in GaAs and Si heterostructures with existing manipulation techniques, enabling quantum error correction using the surface code (cf., e.g., \rcite{fowler2012}). The possibility to tune two-qubit interactions directly using a gate close to the QS makes mediated exchange gates superior to direct exchange gates.

The main findings of this paper are explicit, simple two-qubit gate sequences for STQs, which are mediated by one QS. A single QS can be provided by one QD itself or by another confined system. We also provide expressions for the resulting mediated exchange coupling. The magnetic field gradients are fixed at a constant value and have magnitudes similar to the mediated exchange interactions.\footnote{\label{C7-foot:1}
Magnetic fields are always described in energy units, thus we write the Zeeman Hamiltonian $\frac{g}{2}\mu_B \bm{B}\cdot\bm{\sigma}$ as $\frac{1}{2}\bm{B}\cdot\bm{\sigma}$, where $\mu_B$ is the Bohr magneton and $g$ is the effective g-factor.
} For an empty or a doubly occupied QS, the two-qubit entangling operations via the QS are needed only once if the magnetic field gradients are identical across the DQDs. Such a one-step entangling gate through exchange interactions has never been described before. Two entangling operations together with one single-qubit operation create a controlled $\text{NOT}$ ($\text{CNOT}$) for magnetic field gradients of opposite signs. A singly occupied QS allows a $\text{CNOT}$ operation with two (three) entangling operations with the QS together with single-qubit gates for equal (opposite) magnetic field gradients across the DQDs. These gate sequences realize high-fidelity entangling operations for STQs encoded in GaAs and Si QDs.

The organization of this paper is as follows. \sref{C7-sec:Model} introduces the model that is used for the manipulation of STQs.  The gate sequences that realize entangling operations are constructed in \sref{C7-sec:EntanglingOP}. These sequences differ depending on the occupation of the QS. The gate performances are discussed in \sref{C7-sec:Discussion}. We include limitations from fabrication errors, hyperfine interactions, spin-orbit interactions (SOIs), and charge noise. \sref{C7-sec:CONC} summarizes the results.

%-----------------------------------------------------------------------------------
%-----------------------------------------------------------------------------------
%-----------------------------------------------------------------------------------
\section{Model
\label{C7-sec:Model}}
\begin{figure}[htb]
\centering
\includegraphics[width=0.49\textwidth]{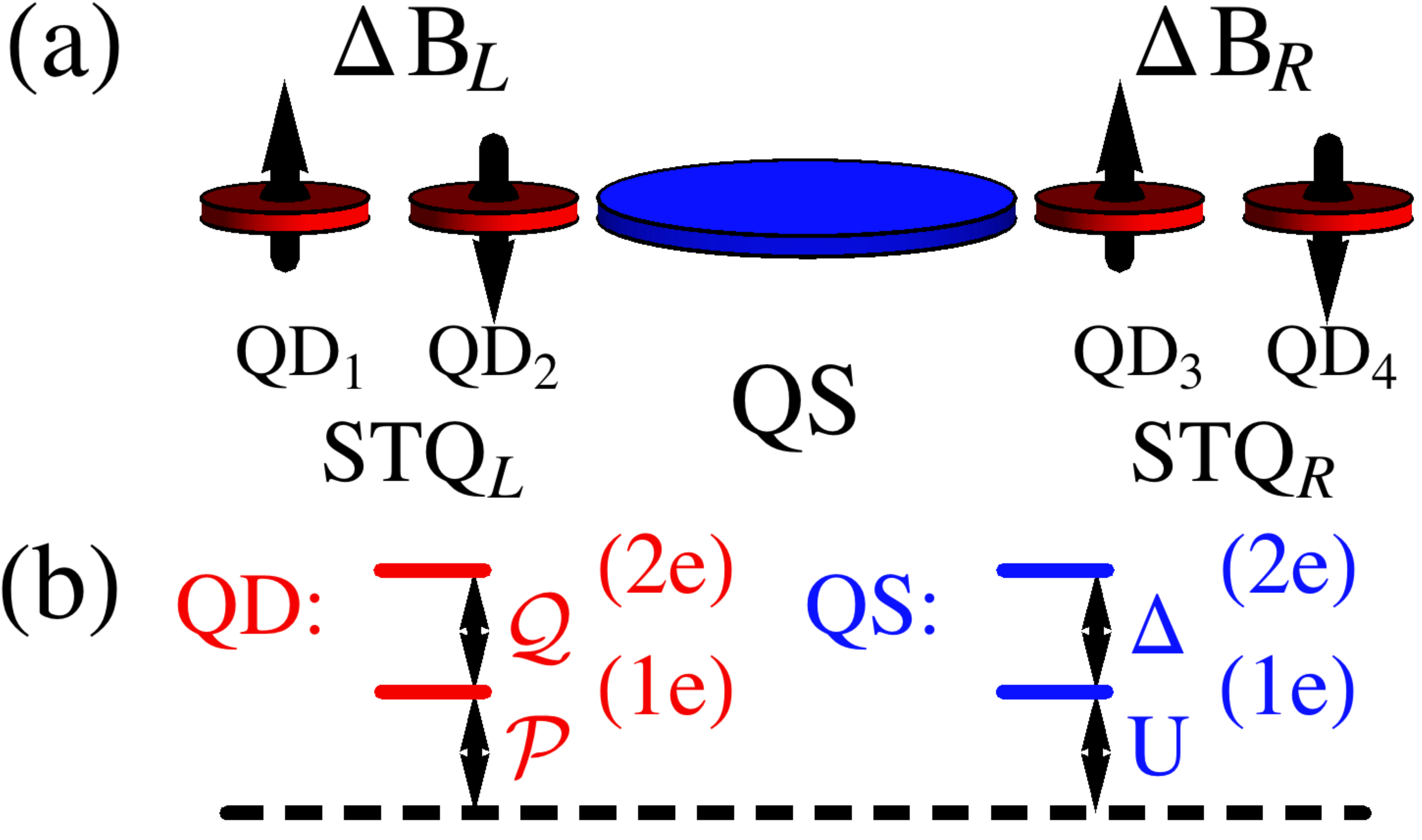}
\caption{Coupling of two STQs via one QS. (a) Four gate-defined QDs, which are shown in red, define two STQs. Each QD is filled with one electron. A global magnetic field acts on all QDs.  There is a small, static magnetic field gradient across the left/right DQD $\Delta B_{L/R}$. We assume identical magnetic field gradients $\Delta B=\Delta B_{L}=\pm\Delta B_{R}$; magnetic fields are equal at the QS and averaged across the DQDs. Exchange interactions together with $\Delta B_{L}$ and $\Delta B_{R}$ are sufficient to control the $s_z=0$ subspace. One QS, which can be provided by another QD, couples $\text{STQ}_L$ and $\text{STQ}_R$. 
(b) Orbital energy levels of the QDs and the QS: adding one electron at the QD requires the energy $\mathcal{P}$, the second electron requires $\mathcal{Q}$. The first electron at the QS costs the energy $U$, and the second electron costs $\Delta$. Adding one electron to the QDs requires the energy $\mathcal{Q}$. The magnitudes of $U$ and $\Delta$ can be tuned using an electric gate close to the QS.
\label{C7-fig:1}}
\end{figure}

We consider an array of four singly occupied QDs ($\text{QD}_1$-$\text{QD}_4$); two QD pairs are coupled by one QS [cf. \fref{C7-fig:1}(a)]. $\text{QD}_1$ and $\text{QD}_2$ encode one STQ, which we call $\text{STQ}_L$ ($\text{QD}_3$ and $\text{QD}_4$ encode $\text{STQ}_R$). A large global magnetic field splits the energies of the ${s_z=0}$ and the ${s_z=\pm 1}$ subspaces of a DQD. We identify the computational subspace with the electron configurations $\left\{\ket{\uparrow\downarrow}_{L,R},\ket{\downarrow\uparrow}_{L,R}\right\}$ on $\text{STQ}_{L,R}$ as the logical qubit states $\left\{\ket{1}_{L,R},\ket{0}_{L,R}\right\}$. The electron configurations $\left\{\ket{\uparrow\uparrow},\ket{\downarrow\downarrow}\right\}$ on the DQDs represent leakage states. Energy $\mathcal{P}$ is needed to fill a QD with one electron, $\mathcal{Q}$ is needed for the second electron. For the QS, energy $U$ is needed to add one electron, and $\Delta$ is needed for a second electron [cf. \fref{C7-fig:1}(b)].

We assume ideal single-qubit gates. In a simplified setting, phase evolutions are generated by the Hamiltonian $\tau_z=\op{1}{1}-\op{0}{0}$; $\tau_x=\op{1}{0}+\op{0}{1}$ creates transitions between the qubit states. A magnetic field gradient $\Delta B_{L}$ over $\text{STQ}_{L}$ 
causes, through
$\frac{\Delta B_{L}}{2}\left(\sigma_{1}^z-\sigma_{2}^z\right)$,\cite{Note1}%\textsuperscript{\ref{C7-foot:1}} 
 a phase evolution $\Delta B_{L} \tau_z^{L}$. 
$\sigma_i^{x,y,z}$ are the Pauli matrices at $\text{QD}_i$. 
Exchange interactions 
$\frac{J_{L}}{4}\left(\bm{\sigma}_{1}\cdot\bm{\sigma}_{2}-\bm{1}\right)$ 
generate qubit rotations $\frac{J_{L}}{2} \tau_x^{L}$. $\bm{\sigma}_i$ is the vector of Pauli matrices on $\text{QD}_i$, $\bm{1}$ is the identity operation, and $J_{L}$ is the exchange coefficient between electrons on $\text{QD}_1$ and $\text{QD}_2$.  
We label exchange gates by $X^{L}_\epsilon=e^{-i 2 \pi \frac{\epsilon}{4}\left[\bm{\sigma}_1\cdot\bm{\sigma}_2-\bm{1}\right]}$, with $\epsilon=\frac{J_{L} t}{h}$, and
phase gates by $Z^{L}_\beta=e^{-i 2 \pi \frac{\beta}{2}\left[\sigma_{1}^z-\sigma_{2}^z\right]}$, with $\beta=\frac{\Delta B_{L} t}{h}$. 
In practice, more complicated gate sequences will likely be needed. As shown in \rcite{cerfontaine2014}, taking relevant experimental details into account, such as finite bandwidth and discrete sampling times, high-fidelity single-qubit gates can indeed be realized with appropriate tuning protocols. The approach taken there could be extended to accommodate such details for our two–qubit gates as well. Equivalent descriptions apply for $\text{STQ}_R$. 
We assume in the whole paper that single-qubit gates are ideal; it is particularly important that independent phase evolutions of $\text{STQ}_L$ and $\text{STQ}_R$ can be realized.

%-----------------------------------------------------------------------------------
%-----------------------------------------------------------------------------------
%-----------------------------------------------------------------------------------
\section{Entangling Operations
\label{C7-sec:EntanglingOP}}

%-----------------------------------------------------------------------------------
%-----------------------------------------------------------------------------------
\subsection{Empty or Doubly Occupied QS}
A nontrivial two-qubit interaction between $\text{STQ}_L$ and $\text{STQ}_R$ can be mediated by an empty or a doubly occupied QS. The configuration with four electrons and an empty QS, which we denote $\left(1,1,0,1,1\right)$, is the ground state if the Fermi energy $E_F$ fulfills 
$E_F\gtrsim 4\mathcal{P}$ and 
$E_F< \left(3\mathcal{P}+U,2\mathcal{P}+U+\Delta,3\mathcal{P}+\mathcal{Q}\right)$. 
The ground-state is $\left(1,1,2,1,1\right)$ with six electrons and a doubly occupied QS if 
$E_F\gtrsim 4\mathcal{P}+U+\Delta$ and 
$E_F < (4\mathcal{P}+U+\mathcal{Q},4\mathcal{P}+2\mathcal{Q},3\mathcal{P}+\mathcal{Q}+U+\Delta)$.

Virtual couplings of the STQs with the QS cause an effective exchange interaction between $\text{QD}_2$ and $\text{QD}_3$:
\begin{align}
	\mathcal{H}_{eff}=\frac{J_{eff}}{4}\left(\bm{\sigma}_2\cdot\bm{\sigma}_3-\bm{1}\right).
	\label{C7-eq:Heff1}
\end{align}
The exchange coefficient $J_{eff}$ can be derived: 
$J_{eff}^0=\frac{4 t^4}{\left(U-\mathcal{P}\right)^2}\left(
\frac{2}{U+\Delta-2\mathcal{P}}+\frac{1}{\mathcal{Q}-\mathcal{P}}
\right)$ 
for an empty QS and 
$J_{eff}^2=\frac{4 t^4}{\left(\mathcal{Q}-\Delta\right)^2}
\left(
\frac{2}{2\mathcal{Q}-\left(U+\Delta\right)}+\frac{1}{\mathcal{Q}-\mathcal{P}}
\right)$ 
for a doubly occupied QS (cf. \aref{C7-ap:EffHam}). 
The tunnel coupling $t$ describes the transfer of electrons between $\text{QD}_{2}$ or $\text{QD}_{3}$ and the $\text{QS}$. 
$t$ is much smaller than any orbital energy differences, which allows us to derive effective low-energy Hamiltonians using Schrieffer-Wolff (SW) perturbation theory \cite{winkler2010,bravyi2011}. Spin effects are relevant in fourth-order SW. Adding two electrons to a quantum level is only permitted in the singlet configuration, making the singlet energy lower. 
We assume that we can tune $J_{eff}$ in \eref{C7-eq:Heff1} to magnitudes similar to $\Delta B_{L/R}$ and restrict $\Delta B=\Delta B_L=\pm\Delta B_R$. The average magnetic fields across each DQD and at the QS are also taken to be identical. The time evolution is described by
\begin{align}
\mathcal{U}_{\epsilon,\beta}^{\pm}=e^{-i 2 \pi \left\{
\frac{\epsilon}{4}\left(\bm{\sigma}_2\cdot\bm{\sigma}_3-\bm{1}\right)+
\frac{\beta}{2}\left(\left[\sigma_1^z-\sigma_2^z\right]\pm\left[\sigma_3^z-\sigma_4^z\right]\right)
\right\}},
\label{C7-eq:UEnt}
\end{align}
with $\beta=\frac{\Delta B t}{h}$, $\epsilon=\frac{J_{eff} t}{h}$.

\begin{figure}[htb]
\centering
\includegraphics[width=0.49\textwidth]{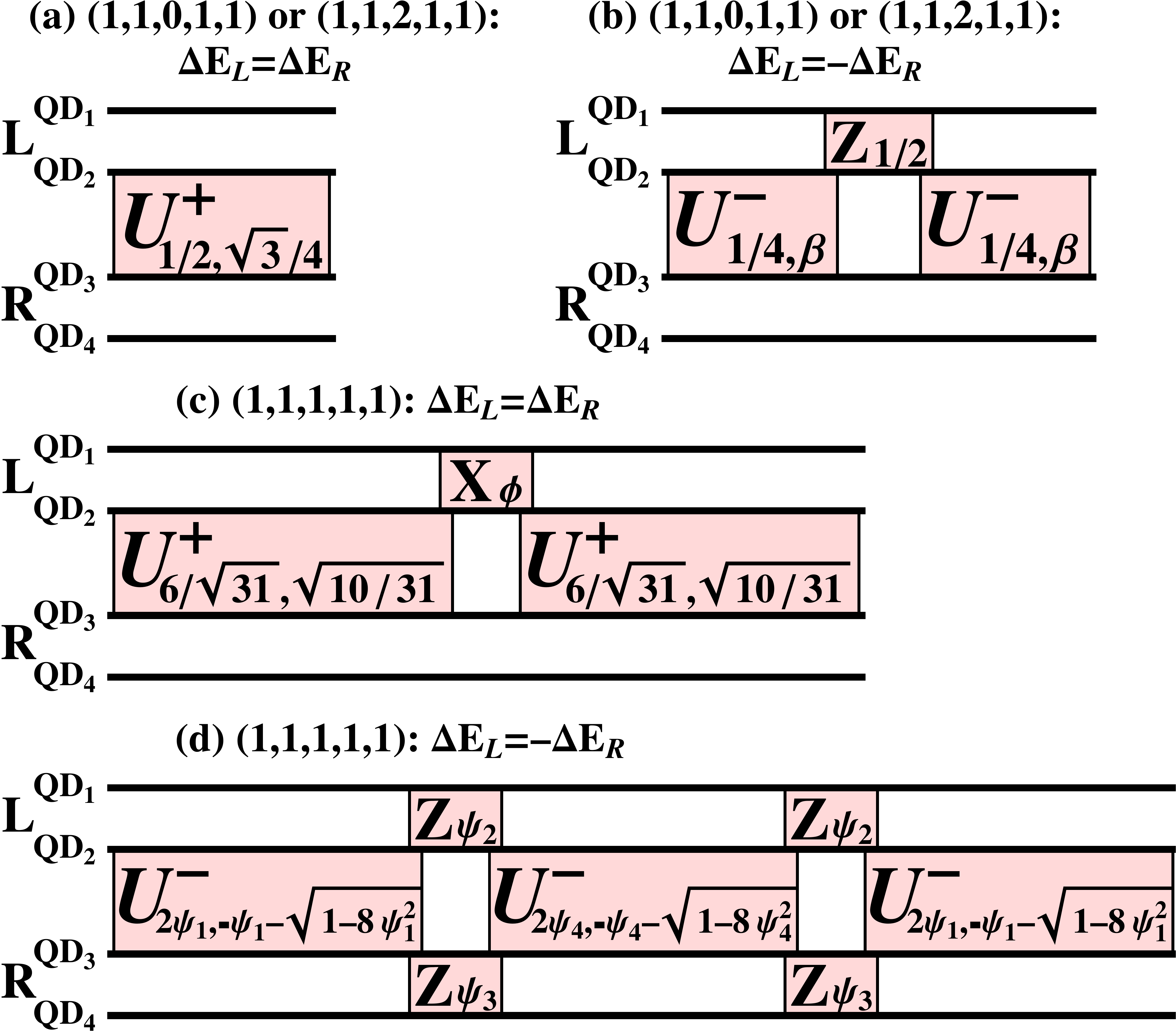}
\caption{Entangling gates that are equivalent to a $\text{CNOT}$ up to single-qubit operations for two STQs coded on $\text{QD}_{1,2}$ and $\text{QD}_{3,4}$. We denote the configurations by the electron numbers at 
($\text{QD}_1$,$\text{QD}_2$,$\text{QS}$,$\text{QD}_3$,$\text{QD}_4$). 
The DQDs are coupled via one QS (cf. \fref{C7-fig:1}). Entangling operations between two STQs mediated by an empty or a doubly occupied QS for (a) equal and (b) opposite magnetic field gradients. The $\text{CNOT}$ operation requires one/two entangling operation according to \eref{C7-eq:UEnt}. Entangling operations mediated by a singly occupied QS for (c) equal and (d) opposite magnetic field gradients. This setup requires two/three entangling operations according to \eref{C7-eq:UEnt2}. All gate sequences and parameters ($\beta$, $\phi$, $\psi_{1-4}$) are discussed in the text.
\label{C7-fig:2}}
\end{figure}

There exists a perfect entangler, which is equivalent to a $\text{CNOT}$ by single-qubit operations, with only one exchange operation for $\Delta B_L=\Delta B_R$: $\mathcal{U}^{+}_{1/2,\sqrt{3}/4}$ [\fref{C7-fig:2}(a)]. Leakage from the computational subspace is absent. One can prove easily that $\mathcal{U}^{+}_{1/2,\sqrt{3}/4}$ is maximally entangling by calculating the Makhlin invariants \cite{makhlin2002} (cf. \aref{C7-ap:FidAnal}). 
The entangling gate uses the exchange operations only once. In previous studies exchange gates were described that needed the exchange interactions twice \cite{levy2002,klinovaja2012}. Even though these studies relate to direct exchange interactions between STQs, our gate can be used without change in these setups. 

The values $\left(\epsilon,\beta\right)=\left(\frac{1}{2},\frac{\sqrt{3}}{4}\right)$ are not the only possible parameters that describe a $\text{CNOT}$. Evaluating $\mathcal{U}^{+}_{\epsilon,\beta}$ from \eref{C7-eq:UEnt} on the ${s_z=0}$ subspace shows that leakage out of the computational subspace is proportional to $\sin\left(2\pi\sqrt{\beta^2+\left(\frac{\epsilon}{2}\right)^2}\right)$: leakage is absent for $2\sqrt{\beta^2+\left(\frac{\epsilon}{2}\right)^2}\in \mathbb{N}$. The Makhlin invariants are $G_1=\cos^2\left(\pi\epsilon\right)$, $G_2=1+2G_1$ under this condition. We obtain a $\text{CNOT}$ operation with $G_1=0$, $G_2=1$ for $\epsilon\in(2\mathbb{N}+1)/2$.

Magnetic field gradients of opposite signs $\Delta B_L=-\Delta B_R$ also permit entangling operations. There is no entangling operation with one coupling to the QS: gates without leakage from the computational subspace have the Makhlin invariants $G_1=1$, $G_2=3$ and are equivalent to single-qubit operations \cite{makhlin2002}. Up to local unitaries, $\text{CNOT}$ is constructed by $
\mathcal{U}^-_{\epsilon,\beta}
Z_{1/2}^{L}
\mathcal{U}^-_{\epsilon,\beta}$, with $\epsilon=(2\mathbb{N}+1)/4$ and finite $\beta$ [\fref{C7-fig:2}(b)]. 
The entangling properties of this sequence are untouched by the value of $\beta$, which means that this operation is independent of the ratio of $\Delta B$ and $J_{eff}$. Levy proposed an equivalent gate sequence for direct exchange interactions between STQs without any magnetic field gradients during the entangling operation \cite{levy2002}.

%-----------------------------------------------------------------------------------
%-----------------------------------------------------------------------------------
%-----------------------------------------------------------------------------------
\subsection{Singly Occupied QS
\label{C7-ssec:SOQL}}
Constructing two-qubit gates for STQs mediated by a singly occupied QS is more challenging because this setup involves more leakage states. The $\left(1,1,1,1,1\right)$ configuration is the ground state for 
$E_F\gtrsim 4\mathcal{P}+U$ and 
$E_F< \left(4\mathcal{P}+\mathcal{Q},3\mathcal{P}+U+\Delta\right)$. 
The mediated interactions between $\text{QD}_2$ and $\text{QD}_3$ can be described by the exchange interactions with the QS:
\begin{align}
	\mathcal{H}_{eff}=\frac{J_{eff}^1}{4}\left[\left(\bm{\sigma}_2\cdot\bm{\sigma}_{QS}-\bm{1}\right)+
	\left(\bm{\sigma}_{QS}\cdot\bm{\sigma}_{3}-\bm{1}\right)\right].
	\label{C7-eq:Heff2}
\end{align}
$J_{eff}^1=2t^2\left(\frac{1}{\mathcal{Q}-U}+\frac{1}{\Delta-\mathcal{P}}\right)$ (cf. \aref{C7-ap:EffHam}) describes direct exchange interactions between $\text{QD}_{2,3}$ and the QS. The couplings between $\text{QD}_{2,3}$ and the QS are identical.
Global magnetic fields are sufficiently strong to consider only one $s_z$ subspace of all five electrons (we choose ${s_z=\frac{1}{2}}$). Besides the computational subspace, which is spanned by $\ket{\uparrow,\downarrow,\uparrow,\downarrow}$, $\ket{\uparrow,\downarrow,\downarrow,\uparrow}$, $\ket{\downarrow,\uparrow,\uparrow,\downarrow}$, $\ket{\downarrow,\uparrow,\downarrow,\uparrow}$ on $\text{QD}_1$-$\text{QD}_4$ coupled to $\ket{\uparrow}$ on the QS, there are six leakage states in the same $s_z$ subspace. We take the magnetic field gradients on $\text{STQ}_{L}$ and $\text{STQ}_{R}$ to be identical $\Delta B=\Delta B_L=\pm\Delta B_R$. Average magnetic fields across each DQD and at the QS are taken to be equal; the time evolution is described by
\begin{align}
	\mathcal{U}^\pm_{\epsilon,\beta}=e^{-i 2 \pi \left\{
\frac{\epsilon}{4}\left[\left(\bm{\sigma}_2\cdot\bm{\sigma}_{QS}-\bm{1}\right)+\left(\bm{\sigma}_{QS}\cdot\bm{\sigma}_{3}-\bm{1}\right)\right]+
\frac{\beta}{2}\left(\left[\sigma_1^z-\sigma_2^z\right]\pm\left[\sigma_3^z-\sigma_4^z\right]\right)
\right\}},
\label{C7-eq:UEnt2}
\end{align}
with $\beta=\frac{\Delta B t}{h}$, $\epsilon=\frac{J_{eff}^1 t}{h}$.

There is an entangling gate for $\Delta B_L=\Delta B_R$ that uses $\mathcal{U}^+_{6/\sqrt{31},\sqrt{10/31}}$ twice together with one single-qubit rotation.
The operation $\mathcal{U}^+_{6/\sqrt{31},\sqrt{10/31}}$ does not cause leakage from the computational subspace and describes the time evolution:
\begin{align}
\left(\begin{array}{cccc}
e^{2 \pi i (4-\sqrt{10})/\sqrt{31}}&0&0&\\
0&e^{8 \pi i/\sqrt{31}}&0&0\\
0&0&1&0\\
0&0&0&e^{2 \pi i (4+\sqrt{10})/\sqrt{31}}
\end{array}\right).
\end{align}
$\mathcal{U}^+_{6/\sqrt{31},\sqrt{10/31}}$ alone is not maximally entangling, as it is described by the Makhlin invariants 
$G_1=\cos^2\left(4\pi/\sqrt{31}\right)\approx 0.40$, $G_2=1+2G_1\approx 1.80$. The sequence 
$\mathcal{U}^+_{6/\sqrt{31},\sqrt{10/31}}
X_{\phi}^L
\mathcal{U}^+_{6/\sqrt{31},\sqrt{10/31}}$ [cf. \fref{C7-fig:2}(c)] has the Makhlin invariants 
$G_1=\big[\cos^2\left(4\pi/\sqrt{31}\right)-\cos\left(2\pi\phi\right)\sin^2\left(4\pi/\sqrt{31}\right)\big]^2$, $G_2=1+2G_1$. 
$\phi=\frac{\arccos\left[\cot^2\left(4\pi/\sqrt{31}\right)\right]}{2\pi}$ 
constructs a gate equivalent to a $\text{CNOT}$; one solution is $\phi\approx0.133001$. 
We did not find any shorter sequences for maximally entangling gates.

We show for completeness also the shortest possible entangling operation that we found if the magnetic field gradients are opposite $\Delta B_L=-\Delta B_R$. A $\text{CNOT}$ operation needs three entangling operations with the QS. Single-qubit phase gates are used between the entangling operations. We get in the notation of \eref{C7-eq:UEnt2}: 
$\mathcal{U}^-_{2\psi_1,-\psi_1-\sqrt{1-8\psi_1^2}}
Z_{\psi_2}^L 
Z_{\psi_3}^R 
\mathcal{U}^-_{2\psi_4,-\psi_4-\sqrt{1-8\psi_4^2}} 
Z_{\psi_2}^L 
Z_{\psi_3}^R$
$\mathcal{U}^-_{2\psi_1,-\psi_1-\sqrt{1-8\psi_1^2}}$
[\fref{C7-fig:2}(d)]. Numerical values for $\psi_{1}-\psi_{4}$ are given \aref{C7-ap:NumVal}.

%-----------------------------------------------------------------------------------
%-----------------------------------------------------------------------------------
%-----------------------------------------------------------------------------------
\section{Gate Performance and Noise Properties
\label{C7-sec:Discussion}}
Entangling two STQs via one QS has advantages compared to direct exchange couplings between STQs. The state energies of the QS are directly tunable using electric gates without affecting the DQDs. It has turned out in experiments that manipulating state energies is easier (cf. especially \rcite{petta2005}) than tuning tunnel couplings \cite{loss1998}. 
Consequently, the setup with a mediating QS also simplifies the realization of entangling operations for weak tunnel couplings $t$. 
Magnitudes of $t$ are on the order of $20~\mu\text{eV}$ and the addition energy $\mathcal{Q}$ reaches a few $\text{meV}$ for single-qubit operations \cite{taylor2007}. Exchange operations are possible with megahertz frequencies: $\nu=(2t^2/\mathcal{Q})/h\approx 200~\text{MHz}$. Reaching large $t$ is very critical for two-qubit gates. DQDs are preferably some distance apart from each other; $t$ decreases exponentially with this distance. One can raise the mediated interaction for small $t$ by significantly lowering $U$ and $\Delta$; the mediated interactions can be completely turned off for large $U$ and $\Delta$. 
It should be possible to raise $J_{eff}$ to magnitudes similar to $\Delta B$. Manipulation frequencies of $200~\text{MHz}$ are sufficient for fast gate operations; experiments with magnetic field gradients with this order of magnitude have been carried out \cite{foletti2009,bluhm2011}. 
Note that two-qubit interactions are tunable independent of the single-qubit parameters.

%------------------------------------------------------------------------------------------------------------------------
%------------------------------------------------------------------------------------------------------------------------
\subsection{Fabrication Errors}
A real system may not fulfill all restrictions of the proposed setup due to fabrication errors:

(1) In our gate constructions, the magnetic field gradients have the same magnitude across the DQDs while only the sign is allowed to differ. The average magnetic field across each DQD is equal to the field at the QS. In reality, only the local magnetic fields at $\text{QD}_{2}$, $\text{QD}_{3}$, and the $\text{QS}$ matter for the proposed gate sequences. $\text{QD}_{1}$ and $\text{QD}_{4}$ are decoupled during the entangling operations. Shifts in their local magnetic fields can be corrected by single-qubit operations. Local magnetic field shifts at the QS are only critical when the QS is singly occupied. In the cases of an empty and a doubly occupied QS, states with an unpaired electron at the QS are only virtually occupied.

(2) The gate construction for the entangling gates assumes that all QDs are identical, especially $\text{QD}_2$ and $\text{QD}_3$ have equal couplings to the QS. The following discussion shows that the gate sequences of \fref{C7-fig:2} permit more general setups, but the robustness against altering the QD parameters depends on the occupation of the QS.

{\emph{Empty/doubly occupied QS --}}  In the cases of an empty QS and a doubly occupied QS, the gate sequences of \fref{C7-fig:2}(a)-(b) can be used if $\text{QD}_2$ and $\text{QD}_3$ differ. \eref{C7-eq:Heff1} remains valid with a modified exchange constant. In fourth-order SW perturbation theory, there is only a modification of the existing exchange term if $\text{QD}_{2}$ differs from $\text{QD}_{3}$:
\begin{align}\nonumber
\widetilde{J}^0_{eff}=
&\sum_{i=1,2}
\frac{2t_1^2t_2^2}{\left(U-\mathcal{P}_i\right)^2\left(\mathcal{Q}_{2i-1}-\mathcal{P}_i\right)}+
\frac{2t_1^2t_2^2}{U+\Delta-\sum_{i=1,2}\mathcal{P}_i}\\
&\times
\left(2\prod_{i={1,2}}\frac{1}{\left(U-\mathcal{P}_i\right)}+\sum_{i={1,2}}\frac{1}{\left(U-\mathcal{P}_i\right)^2}
\right),\\\nonumber
\widetilde{J}^2_{eff}=
&\sum_{i=1,2}
\frac{2t_1^2t_2^2}{\left(\mathcal{Q}_i-\Delta\right)^2\left(\mathcal{Q}_i-\mathcal{P}_{2i-1}\right)}
+\frac{2t_1^2t_2^2}{\sum_{i=1,2}\mathcal{Q}_i-\left(U+\Delta\right)}
\\&\times\left(
2\prod_{i=1,2}\frac{1}{\left(\mathcal{Q}_i-\Delta\right)}+
\sum_{i=1,2}\frac{1}{\left(\mathcal{Q}_i-\Delta\right)^2}
\right).
\end{align}
$t_{1(2)}$ is the tunnel coupling between $\text{QD}_{2(3)}$ and the QS. $\mathcal{P}_{1(2)}$ is the addition energy for an electron to $\text{QD}_{2(3)}$; the second electron costs $\mathcal{Q}_{1(2)}$.

\begin{figure}[htb]
\centering
\includegraphics[width=0.49\textwidth]{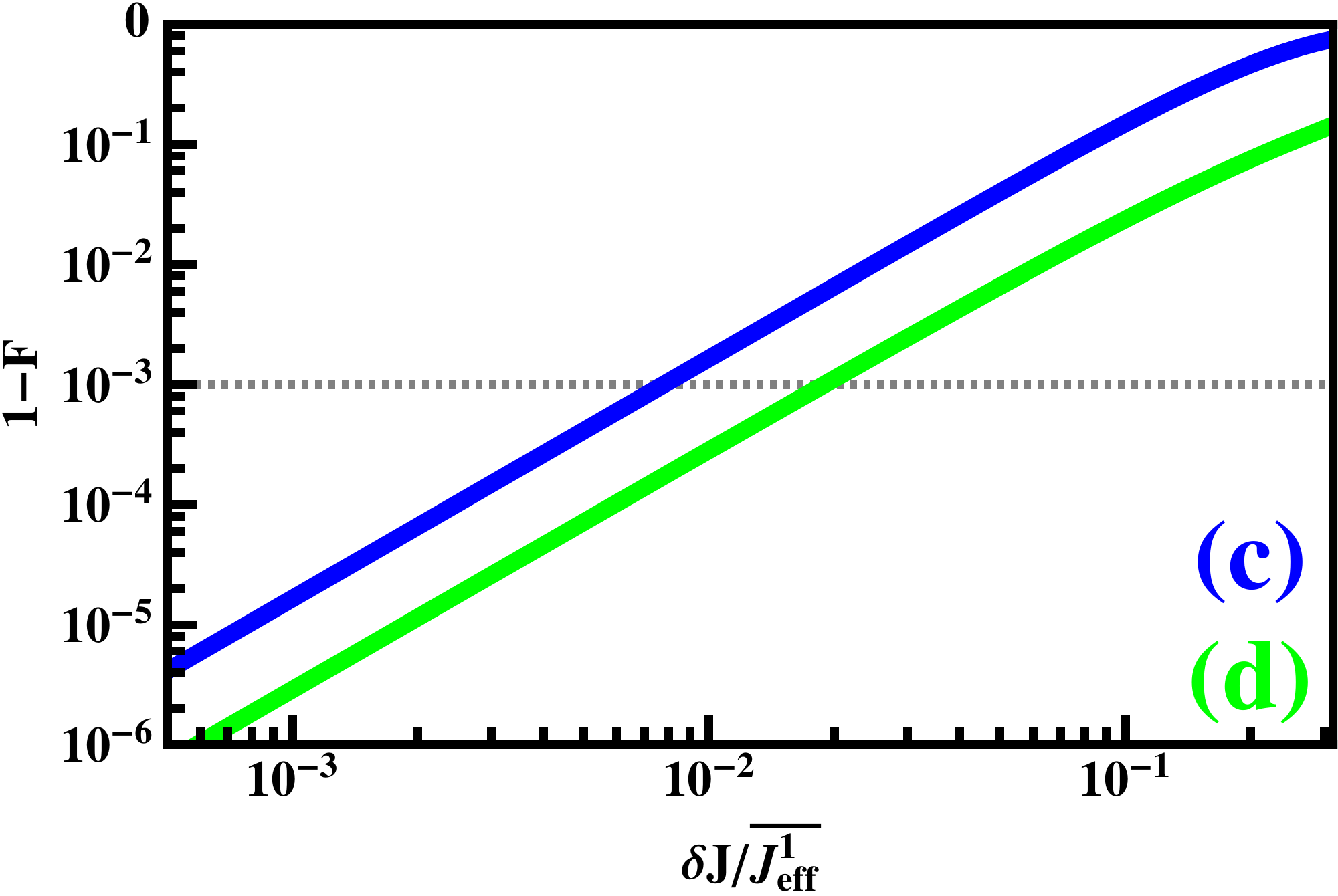}
\caption{Gate infidelities $1-F$ of the entangling gates of \fref{C7-fig:2}(c)-(d) for unequal exchange couplings $J_{eff}^1$ of $\text{QD}_{2}$ with the QS and $\text{QD}_{3}$ with the QS. The difference of the exchange constants $\delta J$ to their average value $\overline{J_{eff}^1}$ is varied in \eref{C7-eq:FabErr}.
\label{C7-fig:3}}
\end{figure}

{\emph{Singly occupied QS --}} In the case of a singly occupied QS, unequal qubit parameters disturb the entangling gates. Differences in the fabrication of $\text{QD}_2$ and $\text{QD}_3$ matter for the entangling operations of \fref{C7-fig:2}(c)-(d). The exchange coupling between $\text{QD}_{2}$ and the QS then differs from $J_{eff}^1$ between $\text{QD}_3$ and the QS. We use instead of \eref{C7-eq:Heff2} a total exchange Hamiltonian:
\begin{align}
\nonumber
\widetilde{\mathcal{H}}_{eff}=&
\frac{
\overline{J_{eff}^{1}}}{4}
\left[
\left(\bm{\sigma}_2\cdot\bm{\sigma}_{QS}-\bm{1}\right)+
\left(\bm{\sigma}_{QS}\cdot\bm{\sigma}_{3}-\bm{1}\right)
\right]
\\\label{C7-eq:FabErr}&+
\frac{
\delta J}{4}\left(\bm{\sigma}_2\cdot\bm{\sigma}_{QS}-\bm{\sigma}_{QS}\cdot\bm{\sigma}_{3}\right),
\end{align}
where $\delta J$ is the difference in the exchange constants and $\overline{J_{eff}^{1}}$ is their average value. \fref{C7-fig:3} shows the gate infidelities as a function of $\delta J/\overline{J_{eff}^1}$. Only strong asymmetries of $\delta J/J_{eff}^1\gtrsim 1\%$ generate gate infidelities of more than $0.1\%$  for the sequences of \fref{C7-fig:2}(c)-(d).

%------------------------------------------------------------------------------------------------------------------------
%------------------------------------------------------------------------------------------------------------------------
\subsection{Hyperfine Interactions}
Hyperfine interactions generate fluctuating magnetic fields locally at the positions of the QDs and the QS. Fluctuations of the nuclear spins are low frequency; they can be treated as static during one entangling operation and only have different distributions for subsequent measurements \cite{coish2005}. A random component $\delta B^z$ parallel to the magnetic field gives the main contribution for strong global magnetic fields. 
For uncorrected nuclear spin baths, typical values for $\delta B^z$ are 
$100~\text{neV}$ ($5~\text{mT}$) in GaAs QDs \cite{taylor2007} and 
$3~\text{neV}$ ($25~\mu\text{T}$) for Si QDs \cite{assali2011}. $\delta B^z$ was suppressed to $10~\text{neV}$ ($0.5~\text{mT}$) in GaAs QDs by preparing the nuclear spin bath in a narrowed state with smaller fluctuations \cite{bluhm2010}. We use these values as the rms of a Gaussian distribution for $\delta B_i^z$ at each QD and at the QS \cite{taylor2007}. We average $1000$ nuclear distributions with a random $\frac{\delta B^z_i}{2}\sigma_z^i$ at each QD and the QS and assume ideal single-qubit gates.

\begin{figure}[htb]
\centering
\includegraphics[width=0.49\textwidth]{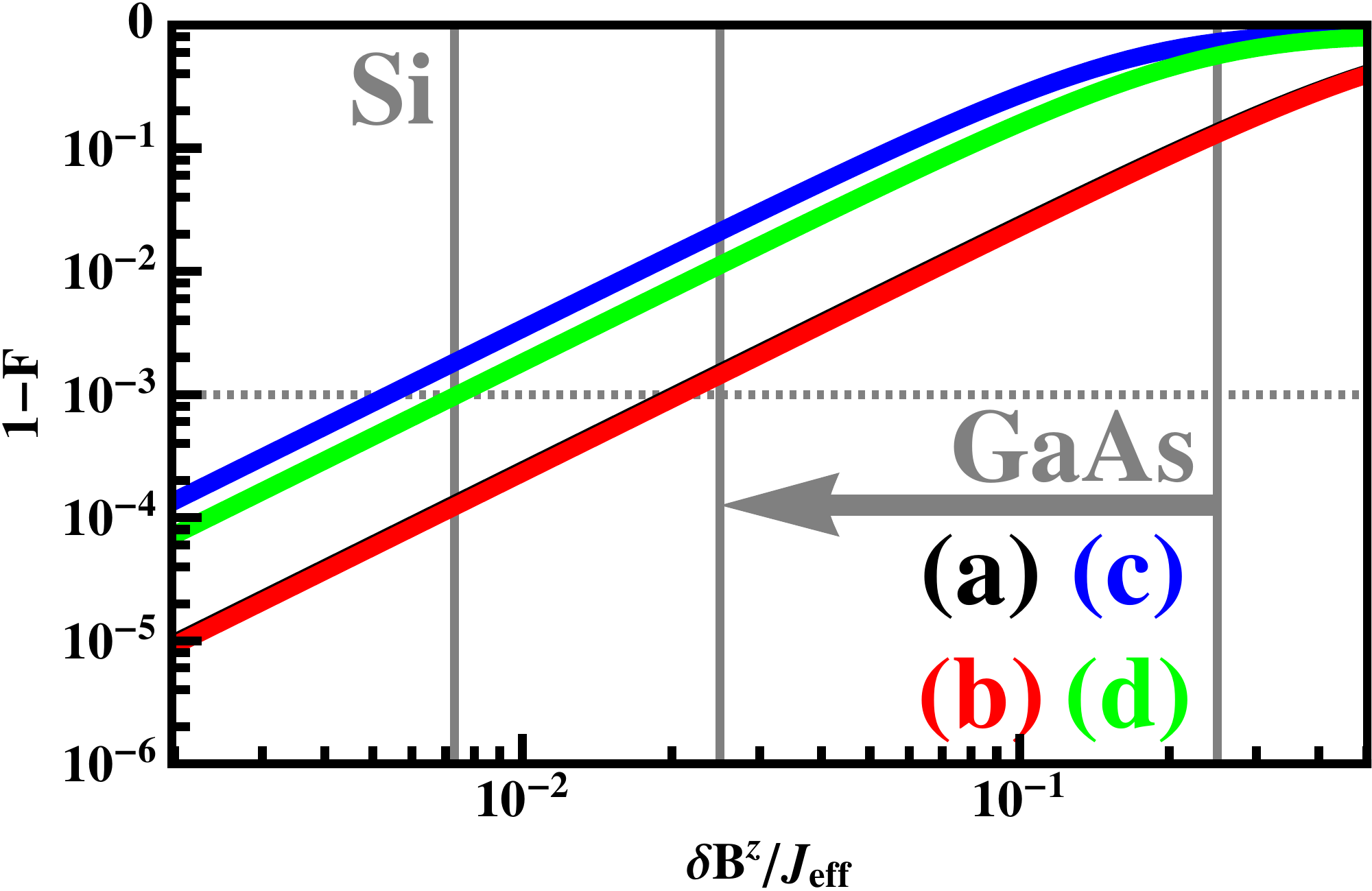}
\caption{Gate infidelities of the entanglement gates of \fref{C7-fig:2}(a)-(d) due to random, local hyperfine fields $\delta B^z_i$ for $J_{eff}/h=200~\text{MHz}$. We vary the ratio of the magnetic field uncertainty $\delta B^z$ and the exchange constant $J_{eff}$. Gray lines mark typical $\delta B^z$ for GaAs and Si QDs. Note that the gate fidelities for GaAs QDs increase strongly when a narrowed distribution of the nuclear spins \cite{bluhm2010} is used instead of an uncorrected spin bath \cite{taylor2007}.
\label{C7-fig:4}}
\end{figure}

\fref{C7-fig:4} shows the gate infidelities $1-F$ of the gate sequences from \fref{C7-fig:2}(a)-(d) as a function of $\delta B^z/J_{eff}$. These gate sequences have infidelities of several percent for GaAs QDs with uncorrected nuclear spin baths, but the errors are suppressed by two orders of magnitude when using a narrowed nuclear spin distribution. One can decrease $\delta B^z$ further by measuring the local hyperfine fields and adjusting the gate sequences in a feedback loop \cite{shulman2014}. All gate sequences reach infidelities of $0.1\%$ for Si QDs. $\delta B^z$ can be suppressed by one order of magnitude in isotopically purified Si compared to natural Si; these heterostructures contain fewer finite-spin nuclei ($^{29}\text{Si}$).

%------------------------------------------------------------------------------------------------------------------------
%------------------------------------------------------------------------------------------------------------------------
\subsection{Spin-Orbit Interactions}
SOIs cause additional errors. The spin rotates slightly when an electron is transferred between localized states. SOIs renormalize the exchange constants weakly. Anisotropic exchange terms introduce errors (cf. \aref{C7-ap:SOI}) \cite{burkard2002,stepanenko2003}. We assume that the magnetic field is oriented in the plane of the QDs, so that the spin-orbit (SO) field is also restricted to this plane. The effective mediated exchange constant is chosen to be $J_{eff}/h=200~\text{MHz}$, and the external global external magnetic field is fixed to $B/h=2.5~\text{GHz}$. This magnetic field strength corresponds to $400~\text{mT}$ in GaAs and $100~\text{mT}$ in Si.
$d\approx 200~\text{nm}$ is a typical distance between localized states. Larger values of $d$ increase the influence of SOIs but decrease the tunnel couplings between localized states. We introduce common SOI parameters \cite{ihn2010,zwanenburg2013}: typical SO lengths are around $l_{so}\approx2~\mu\text{m}$ in GaAs samples. Note that experimentally measured values for $l_{so}$ in GaAs QDs can be much larger \cite{zumbuhl2002,nowack2007} and are strongly probe dependent \cite{ihn2010}. The effective mass in Si heterostructures is nearly three times larger than in GaAs; nanostructures in Si are about two times smaller than in GaAs, while $l_{so}$ is approximately one order of magnitude larger. We use $d= 100~\text{nm}$ and $l_{so}=10~\mu\text{m}$ for Si QDs.

\begin{figure}[htb]
\centering
	\subfigure[$\bm{B}\perp\bm{S}$]{
		\includegraphics[width=0.47\textwidth]{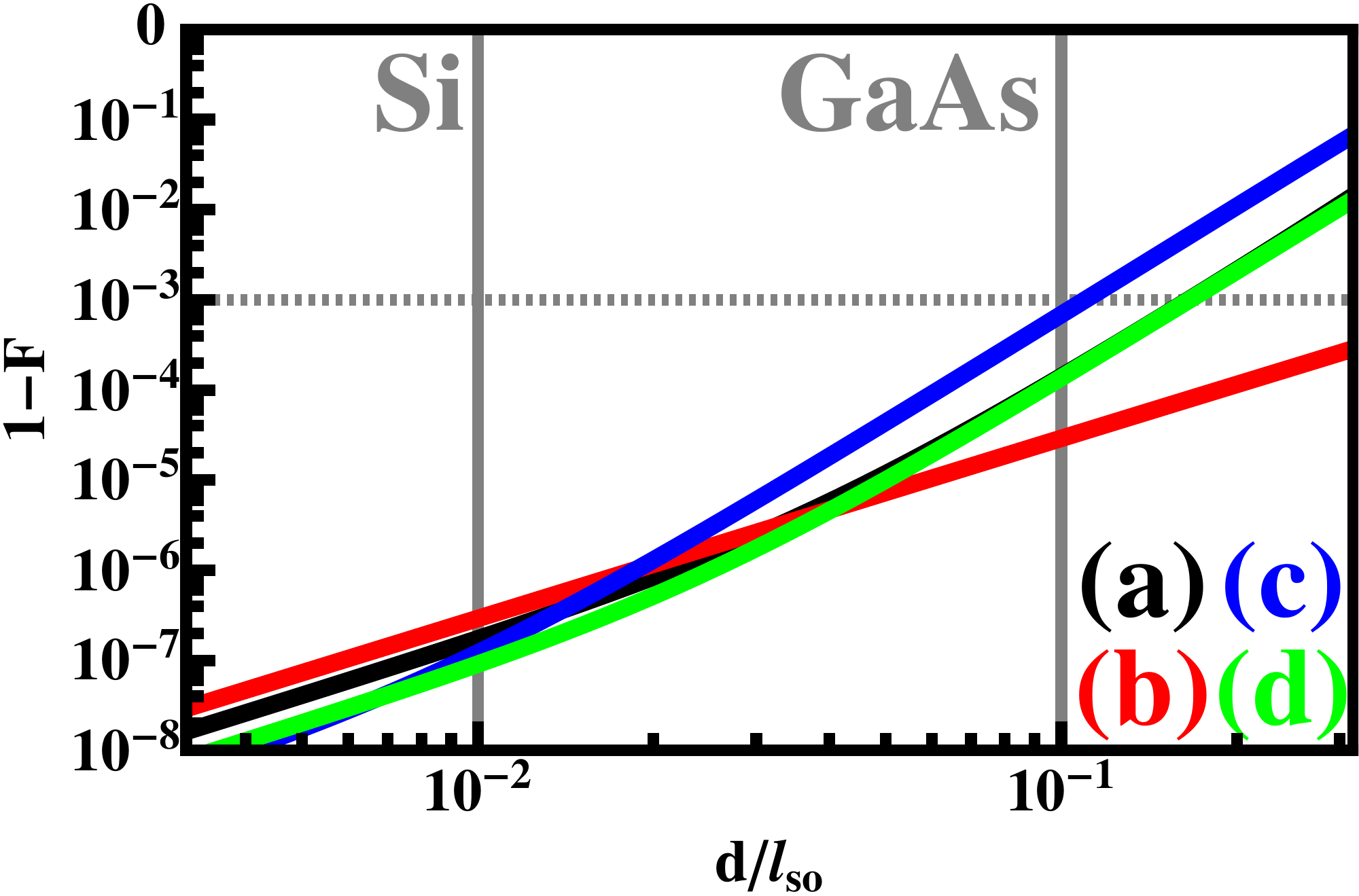}
	}
	\hfill
	\subfigure[$\bm{B}\parallel\bm{S}$]{
		\includegraphics[width=0.47\textwidth]{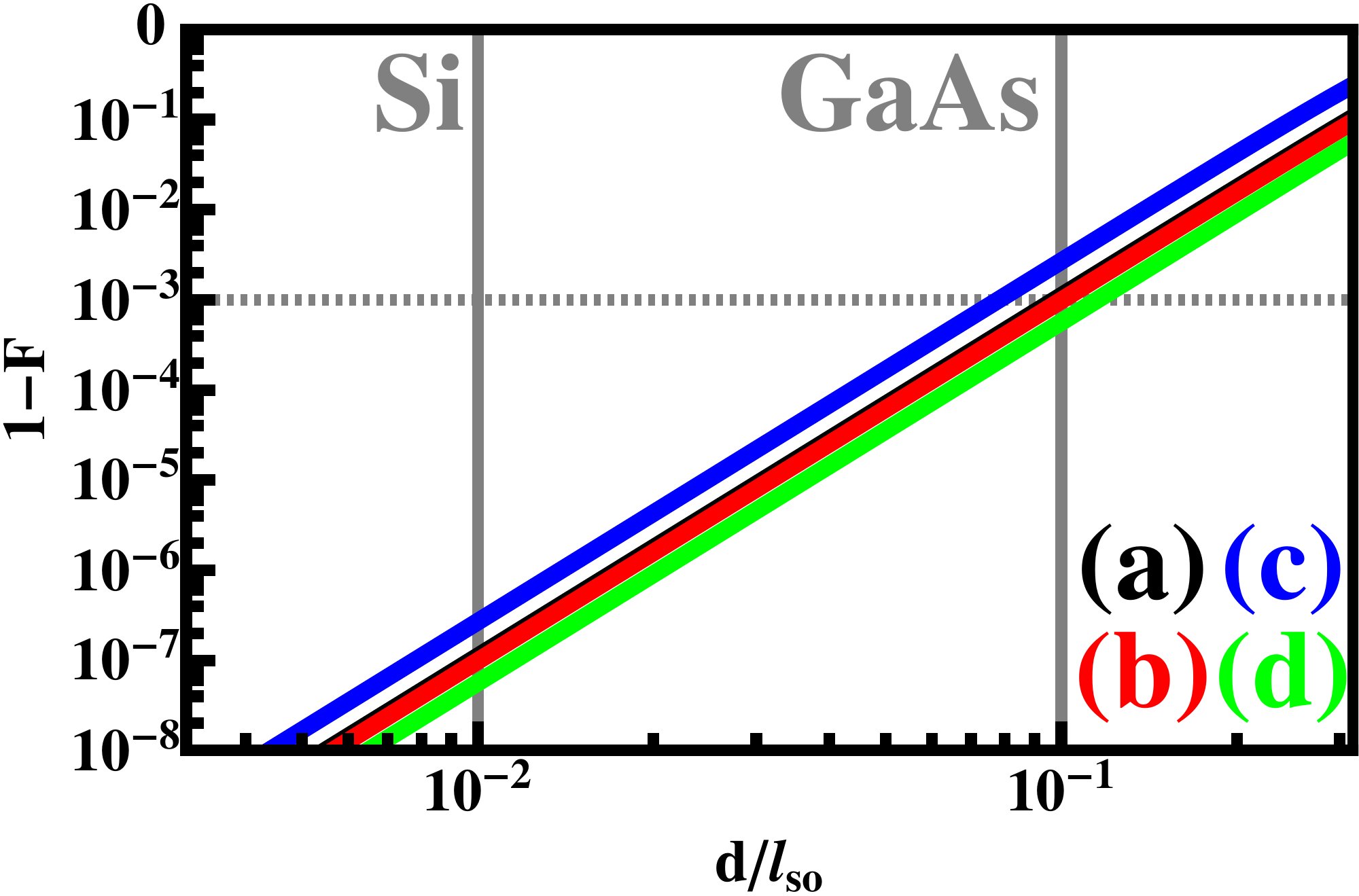}
	}
\caption{Gate infidelities for the gate sequences of \fref{C7-fig:2}(a)-(d) with SOIs for $J_{eff}/h=200~\text{MHz}$. Gray lines mark typical SO parameters for GaAs and Si QDs. $\bm{B}$ describes the external magnetic field, and $\bm{S}$ points along the spin-orbit field (cf. \aref{C7-ap:SOI}). 
\label{C7-fig:5}}
\end{figure}

The gate infidelities $1-F$ for the sequences of \fref{C7-fig:2}(a)-(d) are shown in \fref{C7-fig:5}. 
We assume ideal single-qubit operations. The fidelity analysis shows that SOIs have only a minor effect on the gate sequences. In the worst case, gate infidelities reach a few percent for GaAs QDs. The errors are several orders of magnitudes lower for Si QDs. SOIs are less critical if the external magnetic field is perpendicular to the SO field. In this case, SOIs couple states of different $s_z$, which have a large energy difference \cite{klinovaja2012}. The gate sequences in \rcite{klinovaja2012} were constructed to be optimal with respect to the Dzyaloshinskii-Moriya interaction, which is one part of the anisotropic exchange terms. In any case, our analysis shows that SOIs have only a weak influence on the entangling operations and the gate infidelities hardly increase above $10^{-3}$.

%------------------------------------------------------------------------------------------------------------------------
%------------------------------------------------------------------------------------------------------------------------
\subsection{Charge Noise}
Charge traps of the substrate are uncontrollably filled and unfilled with electrons. These fluctuations, called charge noise, create low-frequency fluctuations of the electric fields at the position of the QDs. We model the dominant effect of charge noise through a zero-frequency fluctuation $\delta \epsilon\left(t\right)$ of the energy difference $\mathcal{C}$ between different charge configurations. $J_{eff}$ is also controlled by $\mathcal{C}$:
\begin{align}
	J_{eff}^{0}\approx J_{eff}^{2}&\approx\frac{4t^4}{\left[\mathcal{C}+\delta \epsilon\left(t\right)\right]^3},\ \ \ 
	J_{eff}^{1}\approx\frac{2t^2}{\mathcal{C}+\delta \epsilon\left(t\right)}.
\end{align}
We disregard, for the case of an empty QS, occupations of states with two electrons at the QS and approximate $\mathcal{C}\approx U-\mathcal{P}\approx \frac{U+\Delta-2\mathcal{P}}{2}$. For a doubly occupied QS, we disregard all states other than in $\left(1,2,1,1,1\right)$, $\left(1,1,1,2,1\right)$, and $\left(1,2,0,2,1\right)$. We approximate $\mathcal{C}\approx \mathcal{Q}-\Delta\approx \frac{2\mathcal{Q}-\left(U+\Delta\right)}{2}$. Charge noise is introduced through the random variable $\delta \epsilon\left(t\right)$ of a Gaussian distribution with rms $\delta \epsilon$; the fidelity is averaged over $1000$ random values of $\delta \epsilon\left(t\right)$. Energy fluctuations in GaAs charge qubits were measured at a few $\mu eV$ ($1~\mu\text{eV}/h\approx 0.24~\text{GHz}$) \cite{petersson2010,dial2013}. Charge noise in Si QDs may be assumed to be of the same order of magnitude.

\begin{figure}[htb]
\centering
	\subfigure[empty/doubly occupied QS]{
		\includegraphics[width=0.47\textwidth]{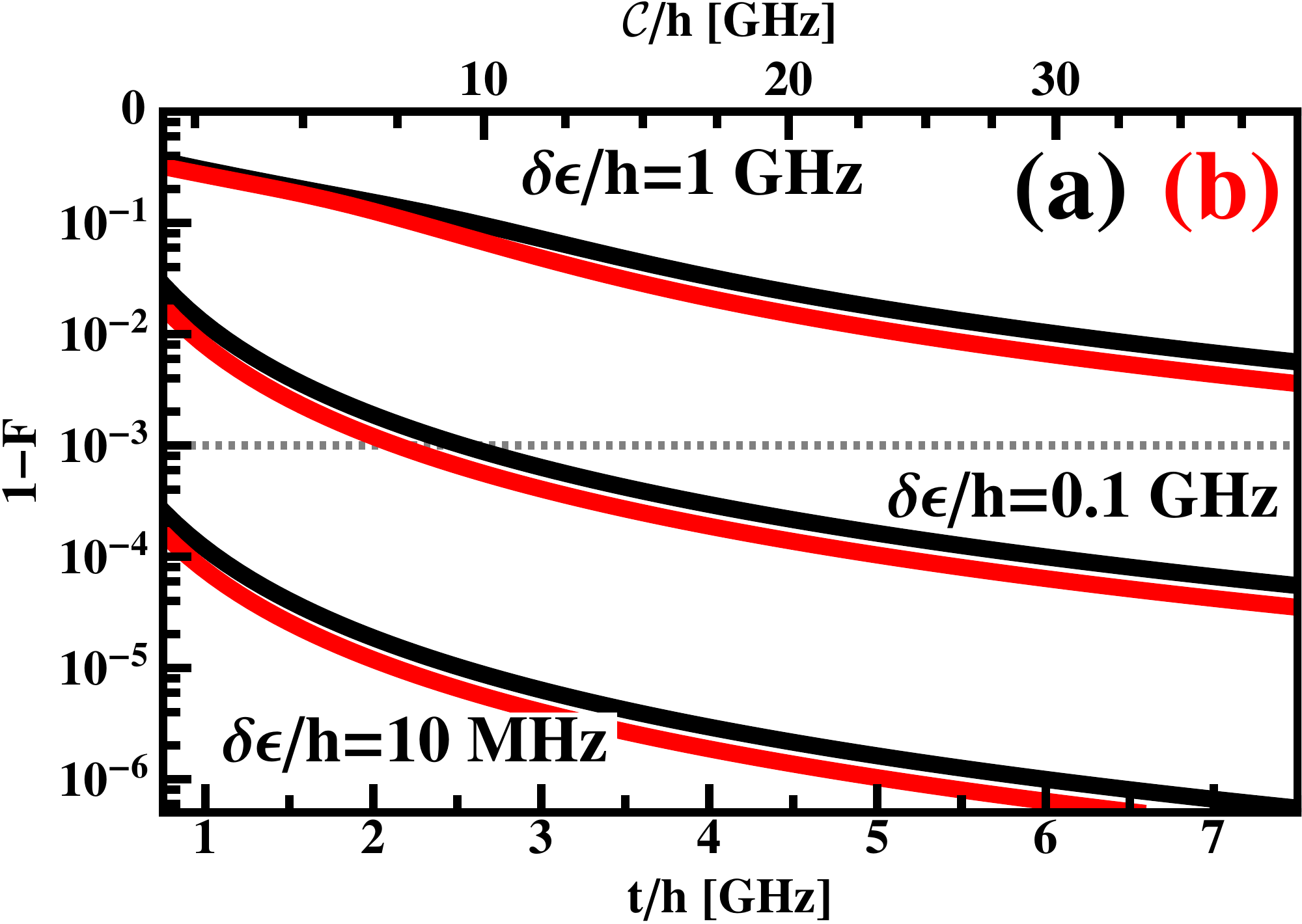}
	}
	\hfill
	\subfigure[singly occupied QS]{
		\includegraphics[width=0.47\textwidth]{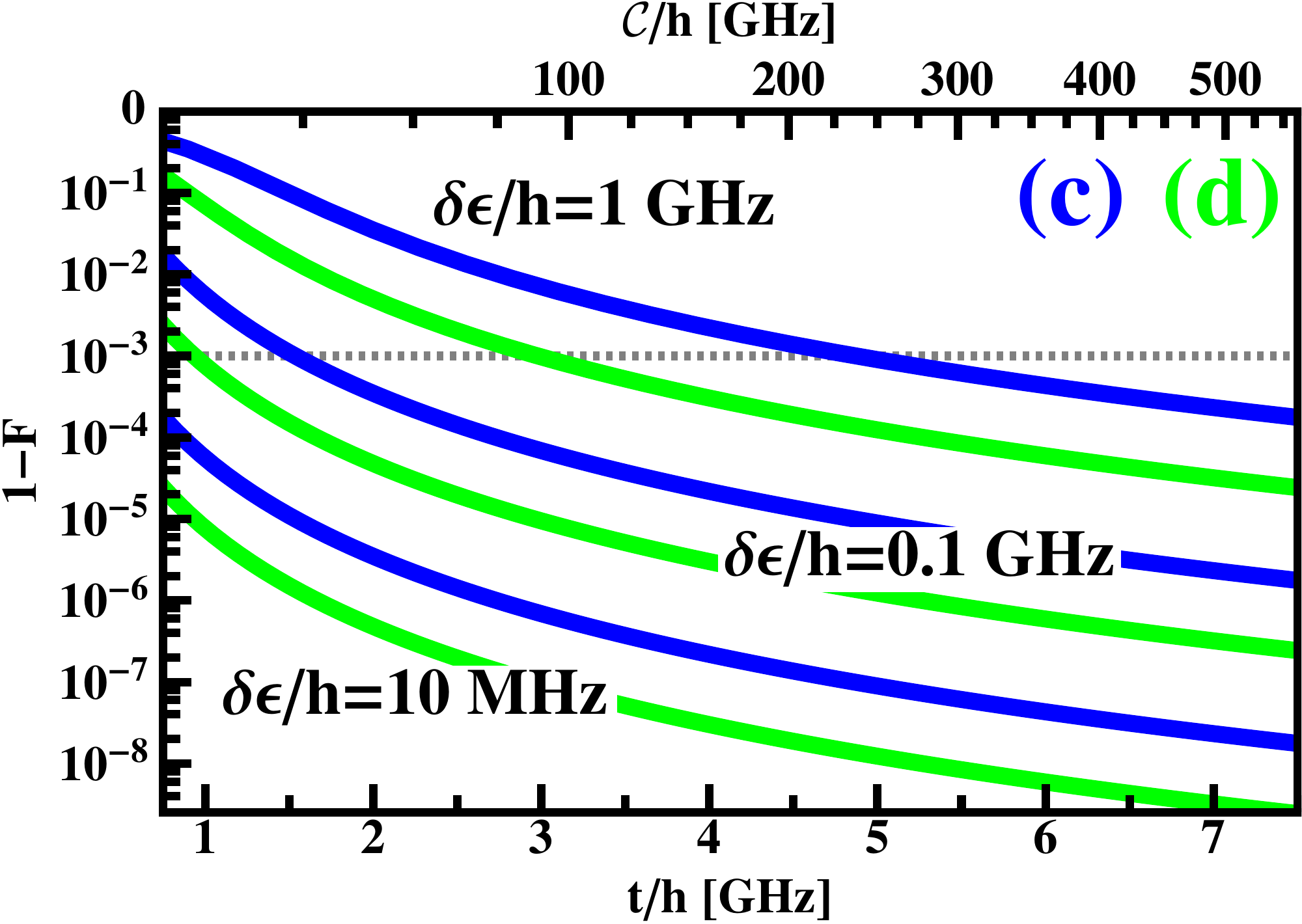}
	}
\caption{Gate infidelity for the gate sequence of \fref{C7-fig:2}(a)-(d) under charge noise $\delta \epsilon$ for $J_{eff}/h=200~\text{MHz}$. Curves for $\delta\epsilon/h=1~\text{GHz}$, $\delta\epsilon/h=0.1~\text{GHz}$, and $\delta\epsilon/h=10~\text{MHz}$ are shown. 
\label{C7-fig:6}}
\end{figure}

\fref{C7-fig:6} shows the influence of charge noise for exchange gates of $J_{eff}/h=200~\text{MHz}$ for ideal single-qubit gates. Charge noise is critical for small $t$. The occupation of energy levels different from the initial charge configuration is higher to reach large $J_{eff}$ for small $t$. Entangling operations via an empty and a doubly occupied QS are more susceptible to charge noise than the operations with a singly occupied QS. $J_{eff}^{0}$ and $J_{eff}^{2}$ require a larger population of the excited energy levels to reach magnitudes similar to $J_{eff}^{1}$. In any case, tunnel couplings of $t/h>3~\text{GHz}$ at $\delta\epsilon/h=0.1~\text{GHz}$ realize entangling operations that have infidelities of less than $0.1\%$.

%-----------------------------------------------------------------------------------
%-----------------------------------------------------------------------------------
%-----------------------------------------------------------------------------------
\section{Conclusion
\label{C7-sec:CONC}}
We have shown that exchange-based entangling operations for two STQs are possible through mediated exchange couplings with one QS. One additional QD or another confined system can provide this QS. The strength of the mediated interactions can be tuned to magnitudes similar to the static magnetic field gradients across the DQDs. It can be controlled independent of the STQs. If the QS is empty or doubly occupied, one needs to use interactions of the QS and the STQs only once if the magnetic field gradients across the DQDs have the same sign. The entangling operations are needed twice for STQs with magnetic field gradients of opposite signs. These gating sequences are also applicable for direct exchange interactions between STQs. A singly occupied QS has slightly lower entangling ability. One needs two operations with the QS if $\Delta B_L$ and $\Delta B_R$ are equal but three if they are opposite to each other. Note that another possibility to couple spin qubits via a mediating QD was proposed recently \cite{srinivasa2013}. However, the entangling mechanism is distinct from our approach; it uses two QSs of a multielectron QD.

Hyperfine interactions introduce major errors if the mediated interactions are of the same size as the uncertainty of the hyperfine fields. Hyperfine interactions can be critical for GaAs QDs; narrowing the nuclear spin distributions for GaAs QDs or choosing Si QDs greatly improves the gate fidelities. Other noise sources and small fabrication errors are less important. In total, optimal gate infidelities of our entangling operations in realistic systems are lower than $10^{-3}$, which is below the threshold of quantum error correction for the surface code \cite{fowler2012}.

Entangling STQs through mediated exchange interactions is very promising, especially since larger arrays of QDs are currently becoming available \cite{braakman2013,braakman2013-2,medford2013,medford2013-2}. Using multielectron QDs for the mediated coupling is also beneficial. The addition energies in these systems are suppressed. Multielectron QDs were successfully explored recently \cite{higginbotham2014}. 
High-fidelity two-qubit gate operations with excellent control should justify the effort of fabricating one QS between the DQDs, rather than coupling them directly.

\headline{Acknowledgments} 
D.D.V. is grateful for support from the Alexander von Humboldt foundation. We acknowledge fruitful discussions with P. Cerfontaine.

%------------------------------------------------------------------------------------
%------------------------------------------------------------------------------------
%------------------------------------------------------------------------------------
\begin{appendix}
%------------------------------------------------------------------------------------
%------------------------------------------------------------------------------------
%------------------------------------------------------------------------------------
\section{Gate Description
\label{C7-ap:FidAnal}}

%------------------------------------------------------------------------------------------------------------------------
%------------------------------------------------------------------------------------------------------------------------
\subsection{Characterization of Entangling Gates}
The Makhlin invariants \cite{makhlin2002,zhang2003} characterize the entangling properties of a gate. The values 
\begin{align}
G_1&=\text{tr}^2\left(m\right)/\left[16~\text{det}(m)\right]\in\mathbb{C}, \\
G_2&=\left[\text{tr}^2(m)-\text{tr}(m^2)\right]/\left[4~\text{det}(m)\right]\in\mathbb{R},
\end{align}
fully characterize two-qubit operations, independent of additional single-qubit operations before and after the gate. $m=M^T_BM_B$, where $M_B$ is the representation of the gate in the Bell-basis. A gate is a perfect entangler if it creates a maximally entangled state from a separable state. It needs to fulfill $\sin^2\left(\gamma\right)\le4\left|G_1\right|\le1$ and $\cos\left(\gamma\right)\left[\cos(\gamma)-G_2\right] \ge 0$ for $G_1=\left|G_1\right|e^{i\gamma}$.  One example is a controlled NOT operation ($\text{CNOT}$), which is characterized by $G_1=0$ and $G_2=1$. We also searched for the square root of a SWAP gate, with $G_1=i/4$ and $G_2=0$. The sequences we found for $\sqrt{\text{SWAP}}$ required more entangling operations with the QS than for the $\text{CNOT}$.

%------------------------------------------------------------------------------------------------------------------------
%------------------------------------------------------------------------------------------------------------------------
\subsection{Fidelity Analysis}
A disturbed operation $U_{d}$ is characterized by the entanglement fidelity \cite{nielsen2000,marinescu2012}: 
\begin{align}
F=\text{tr}
\left[\rho^{RS}
\bm{1}_R\otimes\left(U_{i}^{-1}U_{d}\right)_S
\rho^{RS}
\bm{1}_R\otimes\left(U_{d}^{-1}U_{i}\right)_S
\right].
\label{C7-app:eq:FIDDEF}
\end{align}
$U_{i}$ describes the ideal time evolution. We double the state space to two identical Hilbert spaces $R$ and $S$. $\rho^{RS}=\op{\psi}{\psi}$ represents a maximally entangled state on the larger Hilbert space, e.g., $\ket{\psi}=\left(\ket{0000}+\ket{0110}+\ket{1001}+\ket{1111}\right)/2$. $F$ reaches unity for perfect gates. This definition captures also leakage errors of the qubit.

$U_{d}$ differs from $U_{i}$ through systematic or random errors. We describe random errors with a parameter $\xi$ that modifies $U_{d}\left(\xi\right)$ between different runs of the experiment and obeys a classical probability distribution $f\left(\xi\right)$. The fidelity $F$ is calculated by averaging \eref{C7-app:eq:FIDDEF} over many instances of $U_{d}\left(\xi\right)$ giving $F=\int d\xi~f\left(\xi\right) F\left(\xi\right)$.

%------------------------------------------------------------------------------------------------------------------------
%------------------------------------------------------------------------------------------------------------------------
%------------------------------------------------------------------------------------------------------------------------
\section{Orbital Hamiltonian
\label{C7-ap:EffHam}}
Our description of the system uses the orbital energies of the charge configurations and the transition matrix elements between them. We include in this study $\text{QD}_2$, $\text{QD}_3$, and the QS while considering one orbital at each position (cf. \fref{C7-fig:1}). Each energy level can be empty, singly occupied, or doubly occupied. This treatment corresponds to a Hund-Mulliken approximation \cite{burkard1999}. We describe the electron configurations by the electron numbers on the QDs and the QS: $\left(n_{\text{QD}_2},n_{\text{QS}},n_{\text{QD}_3}\right)$. The electron transfer between the QDs and the QS is described by the spin-conserving hopping Hamiltonian:
\begin{align}
	\mathcal{H}_t=t\sum_{i\in\left\{2,3\right\},\sigma}\left(c^\dagger_{i\sigma}c_{QS\sigma}+\text{H.c.}\right).
	\label{C7-app:eq:HHop}
\end{align}
$c^{\left(\dagger\right)}_{i\sigma}$ is the annihilation (creation) operator of an electron at position $i$ with spin $\sigma$, H.c. is the Hermitian conjugate of the preceding term, and $t\in\mathbb{R}$ is the tunnel coupling.

Adding one electron to a QD requires energy $\mathcal{P}$, and the second electron requires $\mathcal{Q}$. One electron at the $\text{QS}$ requires energy $U$, and a second electron requires $\Delta$ [cf. \fref{C7-fig:1}(b)]. We disregard global magnetic fields as we consider a global $s_z$ subspace in the study of the main text. We assume that energy shifts from local magnetic fields are small compared to the orbital energy scales, especially that the magnetic field gradients across the DQDs fulfill $\Delta B\ll \left(\mathcal{P},\mathcal{Q},U,\Delta\right)$. $\Delta B$ can reach $2.5~\mu\text{eV}$ ($100~\text{mT}$) \cite{foletti2009,brunner2011}, which corresponds to the manipulation frequency $\Delta B/h\approx600~\text{MHz}$ for GaAs nanostructures. Note that the global magnetic field $B$ is large compared to $\Delta B$ [$B=10~\mu\text{eV}$ ($400~\text{mT}$) is a common choice]. The orbital energy scales are usually on the order of a few $\text{meV}$ \cite{taylor2007}. Similar considerations are valid for Si QDs. Note that $\text{QD}_1$ and $\text{QD}_4$ are omitted in the following discussion because they are decoupled during the entangling operations. $\text{QD}_1$ and $\text{QD}_4$ are always singly occupied and add the energies $2\mathcal{P}$ to all electron configurations considered in the main text.

%------------------------------------------------------------------------------------------------------------------------
%------------------------------------------------------------------------------------------------------------------------
\subsection{Empty QS}
The electron configurations can be tuned to $\left(1,0,1\right)$ with an empty QS. The Fermi energy fulfills $E_F\gtrsim 2\mathcal{P}$ and $E_F< \left(\mathcal{P}+U,U+\Delta,\mathcal{P}+\mathcal{Q}\right)$. One can reach the electron configurations $\left(1,1,0\right)$ and $\left(0,1,1\right)$ after one electron transfer. Configurations $\left(2,0,0\right)$, $\left(0,2,0\right)$, and $\left(0,0,2\right)$ are reached after two hopping events. $\mathcal{H}_t$ from \eref{C7-app:eq:HHop} couples states of the same number of spin-up and spin-down electrons on $\text{QD}_{2}$, $\text{QD}_{3}$, and the QS. The problem can be separated into different $s_z$ subspaces $N_{s_z}=N_{\text{QD}_2,\text{QS},\text{QD}_3}^{\uparrow}-N_{\text{QD}_2,\text{QS},\text{QD}_3}^{\downarrow}$ when deriving effective Hamiltonians.

The discussions of the $N_{s_z}=\pm 2$ subspaces are equivalent. We show only the ${N_{s_z}=2}$ subspace. The state notation is fixed to $\ket{\text{QD}_2\uparrow,\text{QD}_2\downarrow,\text{QS}\uparrow,\text{QS}\downarrow,\text{QD}_3\uparrow,\text{QD}_3\downarrow}$. We obtain in the basis $\ket{1,0,0,0,1,0}$, $\ket{1,0,1,0,0,0}$, and $\ket{0,0,1,0,1,0}$ the Hamiltonian:
\begin{align}
\mathcal{H}_{N_{s_z}=2}=\left(
\begin{array}{c|cc}
2\mathcal{P}&-t&-t\\
\hline
-t&\mathcal{P}+U&0\\
-t&0&\mathcal{P}+U\\
\undermat{P}{\textcolor{white}{22\mathcal{P}}}&
\undermat{Q}{\textcolor{white}{\mathcal{P}+U}&\textcolor{white}{\mathcal{P}+U}}
\end{array}
\right).
\end{align}
\smallskip

\noindent
$\mathcal{H}_{N_{s_z}=2}$ provides a perfect example where Schrieffer-Wolff (SW) perturbation theory can be used \cite{bravyi2011,winkler2010}. It describes two energetically separated subspaces, which are weakly coupled. The ground-state subspace $P$ consists of the state $\ket{1,0,0,0,1,0}$. All other states are part of the excited subspace $Q$. The effective Hamiltonian on $P$ in fourth-order SW perturbation theory \cite{winkler2010} describes an energy shift: 
$\text{shift}=-\frac{2 t^2}{U-\mathcal{P}}+\frac{1}{\mathcal{U}-\mathcal{P}}\left(\frac{2 t^2}{U-\mathcal{P}}\right)^2$.

We use the basis $\ket{1,0,0,0,0,1}$, $\ket{0,1,0,0,1,0}$, $\ket{1,0,0,1,0,0}$, $\ket{0,1,1,0,0,0}$, $\ket{0,0,0,1,1,0}$, $\ket{0,0,1,0,0,1}$, $\ket{0,0,1,1,0,0}$, $\ket{1,1,0,0,0,0}$, and $\ket{0,0,0,0,1,1}$ for $N_{s_z}=0$. The total Hamiltonian,
\begin{widetext}
%\begin{small}
\begin{align}
\mathcal{H}_{N_{s_z}=0}=\left(
\begin{array}{cc|ccccccc}
2\mathcal{P} & 0 &	t & 0 & 0 & t &		0 & 0 & 0\\
0 & 2\mathcal{P} &	0 & t & t & 0 &		0 & 0 & 0\\\hline
t & 0 &	\mathcal{P}+U & 0 & 0 & 0 &		t & t & 0\\
0 & t &	0 & \mathcal{P}+U & 0 & 0 &		-t & -t & 0\\
0 & t &	0 & 0 & \mathcal{P}+U & 0 &		-t & 0 & -t\\
t & 0 &	0 & 0 & 0 & \mathcal{P}+U &		t & 0 & t\\
0 & 0 &	t & -t & -t & t &		U+\Delta & 0 & 0\\
0 & 0 &	t & -t & 0 & 0 &		0 & \mathcal{P}+\mathcal{Q} & 0\\
0 & 0 & 0 & 0 & -t & t &	0 & 0 & \mathcal{P}+\mathcal{Q}\\
\undermat{P}{\textcolor{white}{2\mathcal{P}}&\textcolor{white}{2\mathcal{P}}}&
\undermat{Q}{\textcolor{white}{\mathcal{P}+U}&\textcolor{white}{\mathcal{P}+U}&\textcolor{white}{\mathcal{P}+U}\textcolor{white}{\mathcal{P}+U}&\textcolor{white}{\mathcal{P}+U}&\textcolor{white}{U+\Delta}&\textcolor{white}{\mathcal{P}+\mathcal{Q}}&\textcolor{white}{\mathcal{P}+\mathcal{Q}}}
\end{array}
\right),
\end{align}
%\end{small}
\end{widetext}
splits into two weakly coupled subspaces $P$ (at zero energy) and $Q$ (at higher energy). We derive again an effective Hamiltonian on $P$ in fourth-order SW perturbation theory:
\begin{align}
	\widetilde{\mathcal{H}}_P\approx \text{shift}~\bm{1}+\frac{J_{eff}^0}{2}\left(
	\begin{array}{cc}
	-1 & 1\\
	1 & -1
	\end{array}
	\right),
	\label{C7-app:eq:SExch1}
\end{align}
which includes the same energy shift as for $N_{s_z}=\pm 2$. We introduced $J_{eff}^0$ $=$ $\frac{4 t^4}{\left(U-\mathcal{P}\right)^2}$ $\left(\frac{2}{U+\Delta-2\mathcal{P}}+\frac{1}{Q-\mathcal{P}}\right)$.

The total low-energy Hamiltonian on the subspace spanned by the states 
$\ket{1,0,0,0,1,0}$, 
$\ket{1,0,0,0,0,1}$, 
$\ket{0,1,0,0,1,0}$, and
$\ket{0,1,0,0,0,1}$ is
\begin{align}
	\widetilde{\mathcal{H}}_t\approx
	\frac{J_{eff}^0}{4}\left(\bm{\sigma}_2\cdot\bm{\sigma}_3-\mathbf{1}\right).
	\label{C7-app:eq:EX0}
\end{align}
The effective exchange interaction $J_{eff}^0$ lowers only the singlet energy, while it keeps all triplet states untouched. Note that the constant energy shift is neglected in \eref{C7-app:eq:EX0}.

%------------------------------------------------------------------------------------------------------------------------
%------------------------------------------------------------------------------------------------------------------------
\subsection{Singly Occupied QS}
The low-energy subspace of a singly occupied QS consists of the states with the electron configurations $\left(1,1,1\right)$. We reach it for $E_F\gtrsim 2\mathcal{P}+U$ and 
$E_F< (2\mathcal{P}+\mathcal{Q},\mathcal{P}+U+\Delta)$. The interaction between $\text{QD}_2$ and the QS can be separated from the interaction between $\text{QD}_{3}$ and the QS because couplings to excited states are weak. $\mathcal{H}_t$ from \eref{C7-app:eq:HHop} introduces exchange interactions on the low-energy subspace. No couplings are possible for $\left(n_{\text{QD}_2},n_{\text{QS}}\right)=\left(1,1\right)$ in the $\ket{\uparrow,\uparrow}$/$\ket{\downarrow,\downarrow}$ configurations. Singlet pairing lowers the energy of the singlet configuration on $\text{QD}_2$ and QS. $\mathcal{H}_t$ couples to the singlets in $\left(1,1\right)$, $\left(2,0\right)$ and $\left(0,2\right)$. It is straightforward to derive an effective Hamiltonian in second-order SW perturbation theory:
\begin{align}
	\widetilde{\mathcal{H}}_t\approx
	\frac{J_{eff}^{1}}{4}\left(\bm{\sigma}_2\cdot\bm{\sigma}_{QS}-\mathbf{1}\right),
\end{align}
with $J_{eff}^{1}=2t^2\left(\frac{1}{\mathcal{Q}-U}+\frac{1}{\Delta-\mathcal{P}}\right)$. The same result holds for the coupling of the QS to $\text{QD}_3$.

%------------------------------------------------------------------------------------------------------------------------
%------------------------------------------------------------------------------------------------------------------------
\subsection{Doubly Occupied QS
\label{C7-app:EffTwo}}
The last possible case is one doubly occupied QS. The electron configuration $\left(1,2,1\right)$ is the ground state for
$E_F\gtrsim 2\mathcal{P}+U+\Delta$ and 
$E_F < \big(2\mathcal{P}+U+\mathcal{Q},2\left(\mathcal{P}+\mathcal{Q}\right),\mathcal{P}+\mathcal{Q}+U+\Delta\big)$. From the $\left(1,2,1\right)$ configuration, one can reach, with the transfer of one electron, the $\left(2,1,1\right)$ and $\left(1,1,2\right)$ configurations. After a second electron transfer, one can reach the configurations $\left(2,2,0\right)$, $\left(0,2,2\right)$, and $\left(2,0,2\right)$. Deriving an effective Hamiltonian is equivalent to the case of an empty QS. In fourth order SW, we obtain an effective exchange Hamiltonian between $\text{QD}_2$ and $\text{QD}_3$:
\begin{align}
	\widetilde{\mathcal{H}}_t\approx
	\frac{J_{eff}^2}{4}\left(\bm{\sigma}_2\cdot\bm{\sigma}_3-\mathbf{1}\right),
	\label{C7-app:eq:H3}
\end{align}
with $J_{eff}^2=
\frac{4 t^4}{\left(\mathcal{Q}-\Delta\right)^2}
\left(\frac{2}{2\mathcal{Q}-\left(U+\Delta\right)}+\frac{1}{\mathcal{Q}-\mathcal{P}}\right)$. This effect explains the antiferromagnetism of many materials; it is called superexchange in the field of magnetism \cite{anderson1950,mattis2006}.

%------------------------------------------------------------------------------------------------------------------------
%------------------------------------------------------------------------------------------------------------------------
%------------------------------------------------------------------------------------------------------------------------
\section{Spin-Orbit Interactions
\label{C7-ap:SOI}}
SOIs cause spin rotations when an electron moves between localized states. We assume a linear QD arrangement [cf. \fref{C7-fig:1}(a)] and describe the influence of SOIs by \cite{stepanenko2003}
\begin{align}
	\mathcal{H}_{so}=i\mathbf{S} \cdot \sum_{\sigma\sigma^\prime}\left(
	c^\dagger_{2\sigma}\bm{\sigma}_{\sigma\sigma^\prime}c_{QS\sigma^\prime}+
	c^\dagger_{QS\sigma}\bm{\sigma}_{\sigma\sigma^\prime}c_{3\sigma^\prime}
	+\text{H.c.}\right).
	\label{C7-app:eq:HSO}
\end{align}
$\bm{\sigma}=\left(\sigma_x,\sigma_y,\sigma_z\right)^T$ is a vector of Pauli matrices. $i\bm{S}$ describes the transition matrix element between localized states generated by the SOI. It was shown that $\bm{S}$ can be represented by a real vector \cite{stepanenko2012}. $\bm{S}$ defines the direction of the SO field. 
There is a common approximation for localized states which are a distance $d$ apart: $S=\left|\bm{S}\right|\approx t\xi$, with $\xi=\frac{d}{l_{so}}$ and $l_{so}$ being the spin-precession length \cite{stepanenko2012,danon2013,neder2014}. $\xi \ll 1$ for normal GaAs and Si QD pairs.

%------------------------------------------------------------------------------------------------------------------------
%------------------------------------------------------------------------------------------------------------------------
The low-energy Hamiltonian becomes anisotropic when we include, in addition to $\mathcal{H}_t$ in \eref{C7-app:eq:HHop}, the SOIs through $\mathcal{H}_{so}$ from \eref{C7-app:eq:HSO}. We obtain in fourth-order SW perturbation theory additional terms: (1) empty QS,
\begin{align}
	\nonumber
	\widetilde{\mathcal{H}}_{so}^{0}\approx&\frac{1}{\left(U-\mathcal{P}\right)^2}
	\left(\frac{2}{U+\Delta-2\mathcal{P}}+\frac{1}{\mathcal{Q}-\mathcal{P}}\right)
	\\\times
\bigg\{\nonumber&	-S^2\left[
\left(6t^2-S^2\right)\bm{\sigma_2}\cdot\bm{\sigma_3}+\left(2t^2+S^2\right)\bm{1}
\right]\\
&	+4t\left(t^2-S^2\right)\bm{S}\cdot\left(\bm{\sigma}_2\times\bm{\sigma}_3\right)
	+8t^2\left(\bm{S}\cdot\bm{\sigma_2}\right)\left(\bm{S}\cdot\bm{\sigma_3}\right)
	\bigg\},\label{C7-app:eq:SO0}
\end{align}
(2) singly occupied QS,
\begin{align}
\nonumber
	\widetilde{\mathcal{H}}_{so}^1\approx&\left(\frac{1}{\mathcal{Q}-U}+\frac{1}{\Delta-\mathcal{P}}\right)\\\times
	\nonumber
	\bigg\{&	-\frac{S^2}{2}\big[
	\left(\bm{\sigma_2}\cdot\bm{\sigma_{QS}}+\bm{1}\right)+
	\left(\bm{\sigma_{QS}}\cdot\bm{\sigma_{3}}+\bm{1}\right)\big]
	\\\nonumber&
		+t\bm{S}\cdot\big[
\left(\bm{\sigma}_2\times\bm{\sigma}_{QS}\right)+
\left(\bm{\sigma}_{QS}\times\bm{\sigma}_{3}\right)\big]\\
&
	+\left(\bm{S}\cdot\bm{\sigma_2}\right)\left(\bm{S}\cdot\bm{\sigma_{QS}}\right)
	+\left(\bm{S}\cdot\bm{\sigma_{QS}}\right)\left(\bm{S}\cdot\bm{\sigma_{3}}\right)
	\bigg\},
\end{align}
and (3) doubly occupied QS,
\begin{align}
	\nonumber
	\widetilde{\mathcal{H}}_{so}^2\approx&\frac{1}{\left(\mathcal{Q}-\Delta\right)^2}
	\left(\frac{2}{2\mathcal{Q}-\left(U+\Delta\right)}+\frac{1}{\mathcal{Q}-\mathcal{P}}\right)
	\\\times
\bigg\{\nonumber&	-S^2\left[
\left(6t^2-S^2\right)\bm{\sigma_2}\cdot\bm{\sigma_3}+\left(2t^2+S^2\right)\bm{1}
\right]\\
&	+4t\left(t^2-S^2\right)\bm{S}\cdot\left(\bm{\sigma}_2\times\bm{\sigma}_3\right)
	+8t^2\left(\bm{S}\cdot\bm{\sigma_2}\right)\left(\bm{S}\cdot\bm{\sigma_3}\right)
	\bigg\}.\label{C7-app:eq:SO2}
\end{align}

For all charge configurations of the QS, SOIs influence the low-energy subspace similarly. The first term renormalizes the exchange constant. The last two terms describe an anisotropic (super-) exchange interaction. The second term is the dominant contribution, as it scales linearly with $S$ for $S\ll t$. This term is called the Dzyaloshinskii-Moriya interaction in the literature \cite{dzyaloshinsky1958,moriya1960-1,moriya1960-2}. We simplify the expressions in \meref{C7-app:eq:SO0}-\eqref{C7-app:eq:SO2} for $S\ll t$, while we ignore the small renormalization of the exchange constant:
\begin{align}
	\widetilde{\mathcal{H}}_{so}^{0}&\approx
	J_{eff}^{0}\left[
\xi\bm{e}_{\bm{S}}\cdot\left(\bm{\sigma}_2\times\bm{\sigma}_3\right)+
2\xi^2\left(\bm{e}_{\bm{S}}\cdot\bm{\sigma}_2\right)\left(\bm{e}_{\bm{S}}\cdot\bm{\sigma}_3\right)
\right],\\\nonumber
	\widetilde{\mathcal{H}}_{so}^{1}&\approx
	J_{eff}^{1}\bigg\{
\frac{\xi}{2}\bm{e}_{\bm{S}}\cdot\big[
\left(\bm{\sigma}_2\times\bm{\sigma}_{QS}\right)+\left(\bm{\sigma}_{QS}\times\bm{\sigma}_{3}\right)\big]\\
&+
\frac{\xi^2}{2}\big[
\left(\bm{e}_{\bm{S}}\cdot\bm{\sigma}_2\right)\left(\bm{e}_{\bm{S}}\cdot\bm{\sigma}_{QS}\right)+
\left(\bm{e}_{\bm{S}}\cdot\bm{\sigma}_{QS}\right)\left(\bm{e}_{\bm{S}}\cdot\bm{\sigma}_{3}\right)\big]
\bigg\},\\
	\widetilde{\mathcal{H}}_{so}^{2}&\approx
	J_{eff}^{2}\left[
\xi\bm{e}_{\bm{S}}\cdot\left(\bm{\sigma}_2\times\bm{\sigma}_3\right)+
2\xi^2\left(\bm{e}_{\bm{S}}\cdot\bm{\sigma}_2\right)\left(\bm{e}_{\bm{S}}\cdot\bm{\sigma}_3\right)
\right].
\end{align}
$\bm{e}_{\bm{S}}$ is the unit vector pointing along the SO field.

%------------------------------------------------------------------------------------------------------------------------
%------------------------------------------------------------------------------------------------------------------------
%------------------------------------------------------------------------------------------------------------------------
\section{Numerical Gate Search
\label{C7-ap:GaSe}}
We use a numerical gate search algorithm (cf. \rcite{divincenzo2000}), which works similar to the algorithm described by Fong and Wandzura \cite{fong2011}. We define an objective function $f$, that describes the deviation of a gate sequence from an ideal gate. The ideal gate is reached at $f=0$. An example is the construction of a $\text{CNOT}$ on the computational subspace $P$.
The unitary operation on the leakage subspace $Q$ is arbitrary, but the matrix elements between $P$ and $Q$ must vanish. We can search for a $\text{CNOT}$ up to local unitary gates. These gate sequences have the Makhlin invariants $G_1=0$ and $G_2=1$. We construct the objective function $f=\left\|G_1\left(\mathcal{U}_{PP}\right)\right\|+\left\|G_2\left(\mathcal{U}_{PP}\right)-1\right\|+\left\|\mathcal{U}_{PQ}\right\|\ge 0$, where $\left\|\dots\right\|$ describes a matrix norm and $\mathcal{U}_{ij}$ is the projected gate sequence $P_i\mathcal{U}P_j$. $f=0$ for ideal gates.

A gate operation is defined by a sequence of single-qubit operations and two-qubit gates. $X$ and $Z$ rotations, which construct a universal set of single-qubit gates, are characterized by one parameter (cf. description in the main text). The two-qubit gates considered require two parameters. The numerical gate search is constructed in a three step program:

(1) \textit{Initialization --- }
A large number of possible gates is constructed with arbitrary parameters for the single and the two-qubit gates.

(2) \textit{Gate optimization --- }
All gate sequences are optimized. We minimize the objective function $f$. We minimize randomly one, two, or all gates. Most of the time the minimization procedure does not converge.

(3) \textit{Gate selection --- }
We analyze the sequences created in step (2). If the ideal gate is not reached to some accuracy by one gate sequence, we go back to step (2). We keep a collection of gate sequences which are closest to $f=0$ and drop sequences which are far away from the ideal gate.

The obtained gate can usually be simplified. One may especially remove some single-qubit operations from the sequence.

%-----------------------------------------------------------------------------------
%-----------------------------------------------------------------------------------
%-----------------------------------------------------------------------------------
\section{Gate Sequences
\label{C7-ap:NumVal}}

%------------------------------------------------------------------------------------------------------------------------
%------------------------------------------------------------------------------------------------------------------------
\subsection{Full Gate Sequences for CNOT Operations}
We describe the gate sequences to construct a $\text{CNOT}$ operation on the computational subspace in the basis 
$\ket{\uparrow,\downarrow,\uparrow,\downarrow}$, 
$\ket{\uparrow,\downarrow,\downarrow,\uparrow}$, 
$\ket{\downarrow,\uparrow,\uparrow,\downarrow}$, and 
$\ket{\downarrow,\uparrow,\downarrow,\uparrow}$ using one (for an empty/doubly occupied QS) and two (for a singly occupied QS) entangling operations with the QS.

(1) Empty/doubly occupied QS, $\Delta B_L=\Delta B_R$:
\begin{align}
	\text{CPHASE}&=Z_{(3-\sqrt{3})/8}^{L}Z_{(3-\sqrt{3})/8}^R \mathcal{U}^+_{1/2,\sqrt{3}/4},\\
	\text{CNOT}&=\bm{1}\otimes H \times \text{CPHASE} \times	\bm{1}\otimes H,\\
	\bm{1}\otimes H&=X_{1/4}^RZ_{1/8}^RX_{1/4}^R.
\end{align}

(2) Singly occupied QS, $\Delta B_L=\Delta B_R$:
\begin{align}
\text{CNOT}&=\mathcal{U}_{E} \mathcal{U}^+_{6/\sqrt{31},\sqrt{10/31}}X_{\phi}^L\mathcal{U}^+_{6/\sqrt{31},\sqrt{10/31}} \mathcal{U}_{I},\\
\mathcal{U}_{E}&=X_{2\phi_1}^LZ_{\phi_2}^LX_{2\phi_3}^RZ_{1/8}^RX_{1/4}^R
\label{C7-app:eq:1},\\
\mathcal{U}_{I}&=X_{2\phi_4}^LZ_{\phi_5}^LX_{2\phi_6}^LX_{1/4}^RZ_{1/8}^R
\label{C7-app:eq:2}.
\end{align}

%------------------------------------------------------------------------------------------------------------------------
%------------------------------------------------------------------------------------------------------------------------
\subsection{Numerical Values}
The numerical values for the gate sequence of \fref{C7-fig:2}(d) and \meref{C7-app:eq:1}-\eqref{C7-app:eq:2} are:
\begin{align}
	\phi_1&=0.29863890926183401,\\
	\phi_2&=0.39562438490324259,\\
	\phi_3&=0.44782756169938542,\\
	\phi_4&=0.97098194934834639,\\
	\phi_5&=0.30231205192017918,\\
	\phi_6&=0.34055840199539983,
\end{align}
\begin{align}
	\psi_1&=0.25112650148258442,\\
	\psi_2&=0.63771948242765397,\\
	\psi_3&=0.93365278621170444,\\
	\psi_4&=0.22651273139644371.
\end{align}
\end{appendix}
%-----------------------------------------------------------------------------------
%-----------------------------------------------------------------------------------
%-----------------------------------------------------------------------------------

%-----------------------------------------------------------------------------------
%-----------------------------------------------------------------------------------
%-----------------------------------------------------------------------------------
%-----------------------------------------------------------------------------------
\bibliography{library}
\end{document}